\documentclass[aps,prd, showpacs]{revtex4}
\usepackage{graphicx,amsfonts,amssymb,amsbsy}
\usepackage{pstricks}
\usepackage{amsmath}
\usepackage{mathtools}
\begin{document}

\title{Gluon effects on the equation of state of color superconducting strange stars}
\author{E. J. Ferrer}
\affiliation{Department of Physics, University of Texas at El Paso,
El Paso, TX 79968, USA}
\author{V. de la Incera}
\affiliation{Department of Physics, University of Texas at El Paso,
El Paso, TX 79968, USA}
\author{L. Paulucci}
\affiliation{Universidade Federal do ABC, Rua Santa Ad\'elia, 166, 09210-170 Santo Andr\'e, SP, Brazil}

\date{\today}

\begin{abstract}
Compact astrophysical objects are a window for the study of strongly interacting nuclear matter given the conditions in their interiors, which are not reproduced in a laboratory environment. Much has been debated about their composition with possibilities ranging from a simple mixture of mostly protons and neutrons to deconfined quark matter. Recent observations on the mass of two pulsars, PSR J1614-2230 and PSR J0348+0432, have posed a great restriction on their composition, since the equation of state must be hard enough to support masses of about at least two solar masses. The onset of quarks tends to soften the equation of state, but it can get substantially stiffer since in the high-dense medium a repulsive vector interaction channel is opened. Nevertheless, we show that once gluon effects are considered, the equation of state of strange stars formed by quark matter in the color-flavor-locked (CFL) phase of color superconductivity becomes softer decreasing the maximum stellar mass that can be reached. This may indicate that strange stars made entirely of CFL matter can only be favored if other interactions, as the one corresponding to the vector channel, are taken into consideration and are large enough.
\end{abstract}


\maketitle

\section{Introduction}
 Recent very precise mass measurements for two compact objects, PSR J1614-2230 and PSR J0348+0432 with $M=1.97\pm 0.04M_{\odot}$ \cite{Demorest} and $M=2.01\pm 0.04M_{\odot}$ \cite{Antoniadis}, respectively, where $M_{\odot}$ is the solar mass, have provided an important and reliable way to constrain the interior composition of neutron stars. These high mass values imply that the equation of state (EOS) of the corresponding stellar medium should be rather stiff at high densities.

On the other hand, in the highly dense cores of compact objects, the neutron-rich matter can give rise to more degrees of freedom, like hyperons, and perhaps even transitioning to a quark matter phase (see \cite{Lattimer2010} for review). Given that cold strange quark matter has been argued to be absolutely stable \cite{Witten}, a phase transition could occur in the core of a compact star that would quickly favor quark matter in all its interior, thus giving rise to a strange star. 

Although it has been known for more than a decade that perturbative QCD predicts maximum stellar masses for strange stars larger than 2$M_{\odot}$ \cite{Fraga, Kurkela}, there were claims \cite{Trumper, Ozel}, already from the first indications of the existence of 2$M_{\odot}$ stars \cite{Mass-Stars}, that quark matter might have to be ruled out as a possible core phase. One reason was that the density of compact stars is not expected to be sufficiently high to validate a perturbative approach of the strong interaction. On the other hand, nonperturbative calculations based on simple QCD phenomenological models like the NJL with a four-fermion quark-antiquark channel, were found to render a too soft EOS, incapable to stabilize a 2$M_{\odot}$ star against the gravitational collapse (see for example \cite{Bordbar}). However, when other interactions, which are always present in a dense medium of quarks \cite{Kitazawa2002}, such as those corresponding to the diquark channel \cite{Horvath} and/or the vector channel, are taken into account, the EOS can become stiff enough. Because of these findings, quark matter was back in the competition as a possible core phase of massive stars (see for instance \cite{Orsaria}). One may wonder if this last conclusion could be challenged in turn by other effects that might be part of the complicated physics of super dense quark matter.  

Both, weakly coupled QCD and NJL models inspired on the one-gluon exchange interaction of QCD have predicted that the most favored phase of color superconductivity at asymptotically large densities is the color-flavor-locked (CFL) phase \cite{Reviews}. In the case of the NJL approach, the lack of confinement of this model is reflected in the introduction of an undetermined vacuum contribution to the EOS that enters as an independent parameter in the form of a bag constant.  Moreover, in this approach, gluons degrees of freedom are usually disregarded as negligible at zero temperature. Nevertheless, as discussed in the present paper, gluons can affect the EOS of the CFL phase. The reason is connected to the fact that the polarization effects of quarks in the CFL background embed all the gluons with Debye ($m_D$) and Meissner ($m_M$) masses that depend on the chemical potential $\mu$ \cite{Rischke2000}, 
\begin{equation}\label{Masses}
m_D^2=\frac{21-8\ln 2}{18} m_g^2,\;\;\;\;\;\; m_M^2=\frac{21-8\ln 2}{54}m_g^2,\;\;\;\;\;\;m_g^2=g^2\mu^2N_f/6\pi^2.
\end{equation}
As a consequence, the gluons in the CFL phase acquire nonzero rest energies
that yield to a positive contribution to the system's energy density and thus to a negative contribution to the pressure, thereby softening the EOS. In (\ref{Masses}), $N_f$ is the number of flavors of massless quarks, and $g$ the quark-gluon gauge coupling constant.

The main goal of this paper is to investigate the gluon's effects on the EOS and mass-radius (M-R) relationship of quark stars with CFL color superconductivity using a gauged NJL model. We will show that the contribution of the gluon rest energy to the CFL thermodynamic potential softens the EOS and requires a relatively high value of the vector interaction if the observed 2$M_{\odot}$ masses for this class of compact objects is to be reached.

The paper is organized as follows: In Sec. II, the gauged-NJL model is introduced and its one-loop thermodynamic potential in the hard-loop approximation is found. In Sec. III, we investigate the EOS of quark matter in the CFL phase, with and without gluon contributions, for different values of the vector interaction coupling. 
Then, in Sec. IV, the M-R relationships for compact stars having quarks in the CFL phase, with and without gluon contributions, are graphically obtained for a suitable range of parameters and the implications of the results are discussed. In Sec. V, the main conclusions derived from the paper outcomes are listed.

\section{Gauged-NJL model with vector interactions}

Consider the three-flavor gauged-NJL Lagrangian at finite baryon density, which is invariant under $SU_c(3)\times SU_L(3)\times SU_R(3)\times U_V(1)$ symmetry
\begin{eqnarray} \label{Lagrangian}
  \mathcal{L}=-\bar{\psi}(\gamma^\mu D_\mu+\mu\gamma^0)\psi-G_V(\bar{\psi}\gamma_\mu\psi)^2+G_S\sum_{k=0}^8\left [ \left (\bar{\psi} \lambda_k \psi\right )^2 +\left (\bar{\psi}i\gamma_5\lambda_k \psi \right )^2 \right ]  \qquad\nonumber
\\
-K\left [det_f \left ( \bar{\psi}(1 +\gamma_5)\psi \right ) +det_f \left ( \bar{\psi}(1 +\gamma_5)\psi \right )  \right ]
  +\frac{G_D}{4}\sum_\eta(\bar{\psi}P_\eta\bar{\psi}^T)(\psi^TP_\eta\psi)+\mathcal{L}_G,
\end{eqnarray}
In (\ref{Lagrangian}), the quark fields $\psi_i^a$ have flavor ($i={u,d,s}$) and color ($a={r,g,b}$) indexes. We consider the quark-antiquark channel with coupling constant $G_S$, the 't Hooft determinant term that excludes the $U_A(1)$ symmetry \cite{'tHooft} with coupling constant $K$, the diquark channel $P_\eta=C\gamma_5\epsilon^{ab\eta}\epsilon_{ij\eta}$ with coupling $G_D$, and the repulsive vector channel with coupling $G_V$. This last interaction naturally appears after a Fierz transformation of a point-like four-fermion interaction with the Lorentz symmetry broken by the finite density \cite{Buballa}. We neglect the current quark masses $m_{0i}$ since our main interest is to explore the stable CFL phase, which can only exist in the density region where $m_s^2<2\mu \Delta_{CFL}$. In this way, we avoid the chromomagnetic instabilities of the gCFL phase \cite{CI-gCFL}. Thus, our results are valid in the density domain where the contribution $\mu^2\Delta_{CFL}^2$ to the EOS is more important than that of the s-quark mass. Hence, it is consistent to neglect $m_s$ in our calculations. To open the door to possible $m_s$ effects relevant at lower densities, the stable ground state that removes the chromomagnetic instability at those densities would have to be determined first, a problem out of the scope of this paper. Despite several propositions \cite{CFL-Loff}-\cite{Gluon-C}, there is still not a definitive accepted answer.

The gluon Lagrangian density $\mathcal{L}_G$ is
\begin{equation} \label{Lagrangian-g} 
  \mathcal{L}_G=-\frac{1}{4}G_{\mu\nu}^AG^{\mu\nu}_A +\mathcal{L}_{gauge}+\mathcal{L}_{ghost},
\end{equation}
with $G_{\mu\nu}^A=\partial_\mu G^A_\nu-\partial_\nu G^A_\mu +gf^{ABC}G^{B}_\mu G^{C}_\nu$, the gluon strength tensor;
$\mathcal{L}_{gauge}$, a gauge-fixing term; and $\mathcal{L}_{ghost}=-\eta^{A\dagger}\partial^{\mu}(\partial_{\mu}\eta^{A}+gf^{ABC}G^{B}_{\mu}\eta^{C})$, the ghost-field Lagrangian.
The coupling between gluons and quarks occurs through the covariant derivative 
\begin{equation} \label{CovDerivative}
D_\mu= \partial_\mu -i gT^AG^A_\mu,
\end{equation}
where $T^A=\lambda^A/2$, $A=1-8$,  are the generators of the $SU_c(3)$ group in the fundamental representation ($\lambda^A$ are the Gell-Mann matrices). The model (\ref{Lagrangian}) is nonrenormalizable due to the four-fermion interaction terms, so a cutoff $\Lambda$ needs to be introduced in all the calculations to regularize the theory in the ultraviolet region. The parameter $\Lambda$ defines the energy scale below which this effective theory is valid.  

We are interested in exploring the region of densities large enough to have a CFL phase. Then, we can use that the chiral condensate has already vanished and hence consider only the expectation values for the diquark condensate $\Delta_\eta=\langle\psi^TP_\eta\psi\rangle$ and the baryon charge density $\rho=\langle\bar{\psi}\gamma_0\psi\rangle$. One can now bozonize the four-fermion interaction via a Hubbard-Stratonovich transformation and then take the mean-field approximation to obtain the mean-field Lagrangian 

\begin{equation} \label{MF-Lagrangian}
  \mathcal{L}_{MF}=-\bar{\psi}(\gamma^\mu D_\mu+\widetilde{\mu}\gamma^0)\psi+\frac{1 }{2}(\bar{\psi}P_\eta\bar{\psi}^T)\Delta_\eta+\frac{1}{2}\Delta^*_\eta(\psi^TP_\eta\psi)
  -\frac{\Delta_\eta\Delta^*_\eta}{G_D}+G_V \rho^2 + \mathcal{L}_G,
\end{equation}
where $\widetilde{\mu}=\mu-2G_V \rho$. 

The partition function in the mean-field approximation is then given by
\begin{equation} \label{MF-PartitionF}
  Z_{MF}=\mathcal{N}\int \mathcal{D}\bar{\psi}\mathcal{D}\psi \mathcal{D}G \mathcal{D}\eta^{\dagger} \mathcal{D}\eta e^{ \int \mathcal{L}_{MF}d^{4}x} 
\end{equation}

Even though the mean-field Lagrangian (\ref{MF-Lagrangian}) is quadratic in the fermion fields, the presence of fermion-fermion terms does not allow to use the conventional formula to integrate in the fermion fields.  However, we can use the well-known Nambu-Gorkov (NG) method to overcome this difficulty. For that, we rewrite $\mathcal{L}_{MF}$ in terms of $\psi$ and its charge conjugate $\psi_C=C\bar{\psi}^{T}$, with $C=i\gamma^2\gamma_0$  the charge operator, and introduce NG spinors 
\[\Psi=\begin{pmatrix}\psi\\
\psi_C\end{pmatrix}\]

Now we can transform to Euclidean variables and consider a space with volume $V$ and Euclidean time of length $\beta=1/T$, to obtain the finite-temperature, mean-field partition function  
\begin{equation} \label{MF-Partition-T}
  Z_{MF\beta}=e^{-\beta V\Omega_{MF}}=\mathcal{N}e^{-\beta V(- \frac{\Delta_\eta\Delta^*_\eta}{G_D}+G_V \rho^2)}\int \mathcal{D}\bar{\Psi}\mathcal{D}\Psi \mathcal{D}G \mathcal{D}\eta^{\dagger} \mathcal{D}\eta e^{ I_{NG}+\int \mathcal{L}_{G}}
\end{equation}
where
\begin{equation} \label{MFaction}
 I_{NG} = \frac{1}{2}\int\bar{\Psi}{\cal S}_{G}^{-1}\Psi 
\end{equation}
is the NG action with 
\begin{equation}
{\cal S}_{G}^{-1}(x)=\left(
\begin{array}{cc}
\gamma^\mu D_\mu+\widetilde{\mu}\gamma^0 & \Phi^{-}\\
\Phi^+ & \gamma^\mu D_\mu-\widetilde{\mu}\gamma^0
\end{array}\right),
\label{fullpropG}
\end{equation}
the inverse NG fermion propagator in the color superconducting medium. Notice that ${\cal S}_{G}^{-1}(x)$ depends on the gluon field through the covariant derivative. The symbol $\Phi^+=\Delta_\eta P_\eta$ denotes the CFL gap matrix and $\Phi^-=\gamma_0[\Phi^+]^\dagger\gamma_0$ \cite{Pisarski}.  Replacing in (\ref{fullpropG}) the covariant derivative by the derivative implies ${\cal S}_{G}^{-1}(x)\to {\cal S}_{G}^{-1}(x)|_{G=0}$, where ${\cal S}_{G}^{-1}(x)|_{G=0}$ is the mean-field inverse fermion propagator of the CFL phase of the non-gauged NJL model. 

Performing the integral in the NG spinors in (\ref{MF-PartitionF}) we obtain
\begin{equation} \label{GeffectiveZ}
  Z_{MF}=e^{-\beta V\Omega_{MF}}=\mathcal{N}e^{-\beta V(- \frac{\Delta_\eta\Delta^*_\eta}{G_D}+G_V \rho^2)}\int \mathcal{D}G^A_{\mu} \mathcal{D}\eta^{\dagger} \mathcal{D}\eta e^{-\Gamma(\mathbf{G})+\int \mathcal{L}_Gd^4x },
\end{equation}
where 
\begin{equation} \label{det}
 e^{-\Gamma(\mathbf{G})}=[\beta \mathrm{Det} {\cal S}_{G}^{-1}]^{\frac{1}{2}}
\end{equation} 

The effective action $\Gamma(\mathbf{G})$ can be expanded in powers of the gluon fields as
\begin{equation} \label{GammaExpansion}
  \Gamma(\mathbf{G})=\beta V\Omega_0+ \sum_{n=2}^{\infty}\int dx_{1}...dx_{n}\Pi^{A_1,A_2,...,A_n}_{\mu_1,\mu_2,...\mu_n}(x_1,x_2,...x_n)G_{A_1}^{\mu_1}(x_1)...G_{A_n}^{\mu_n}(x_n)
\end{equation} 
where the tensors $\Pi^{A_1,A_2,...,A_n}_{\mu_1,\mu_2,...\mu_n}$ are n-vertex (with vertex $gT^A\gamma_\mu$) one-loop polarization operators with internal lines of fermions with propagator
\begin{equation}
{\cal S}_G(x)|_{G=0}=\left(
\begin{array}{cc}
S^+(x) & \Xi^-(x)\\
\Xi^+(x) & S^-(x)
\end{array}\right).
\label{fullprop}
\end{equation}
Here
\begin{equation}\label{Inv-Prp}
S^{\pm}=([S_0^{\pm}]^{-1}-\Sigma^{\pm})^{-1}, 
\end{equation}
denotes the propagators for quasiparticles and charge-conjugate quasiparticles, with $S_0^\pm$ 
the propagator for massless free quarks, and 
\begin{equation}
\Sigma^{\pm}=\Phi^{\mp}S_0^{\mp}\Phi^\pm,
\end{equation}\label{Inv-Prp} 
\\
the quark self-energy generated by the exchange of particles or conjugate particles with the color superconducting condensate. The off-diagonal elements of (\ref{fullprop}) are given by
\begin{equation}
\Xi^{\pm}=-S_0^{\mp}\Phi^{\pm}S^{\pm}. 
\end{equation}\label{Off-Diag}

Except for the replacement of $\mu$ by $\widetilde{\mu}$, due to the vector interaction, the zero-order term $\Omega_0$ in (\ref{GammaExpansion}) exactly reproduces the fermion determinant part of the conventional (no gluons and no vector interaction) mean-field thermodynamic potential of the CFL-phase \cite{CFL}. Going to momentum space and summing in the Matsubara frequencies, the quark contribution to the thermodynamic potential can be written as
\begin{equation} \label{Gamma0}
\Omega_{CFL}(T)= \Omega_0+\frac{3\Delta^2}{G_D}-G_V\rho^2 = \-\frac{-1}{4\pi^2}\int_0^\Lambda dp p^2 \sum_{i}[ |
\epsilon_i|+\frac{2}{\beta} \ln(1+e^{-\beta |
\epsilon_i|} ]+\frac{3\Delta^2}{G_D}-G_V\rho^2, 
\end{equation} 
where the $\epsilon_i$'s are the dispersion relations of the quasi-quarks/antiquarks in the CFL phase and we used that in the chiral limit $\Delta_1=\Delta_2=\Delta_3=\Delta$.

In the zero-temperature limit the logarithm vanishes and we obtain 
\begin{equation} \label{Gamma0}
\Omega_{q}= \Omega_{CFL}(T=0)= -\frac{1}{4\pi^2}\int_0^\Lambda dp p^2 (16|
\epsilon|+16|\overline{\epsilon}|)-\frac{1}{4\pi^2}\int_0^\Lambda
dp p^2 (2|\epsilon'|+2|\overline{\epsilon'}|) +\frac{3\Delta^2}{G_D}-G_V\rho^2 ,
\end{equation} 
with
\begin{equation}\label{Spectra}
\varepsilon=\pm \sqrt{(p-\tilde{\mu})^2+\Delta^2}, \quad
\overline{\varepsilon}=\pm \sqrt{(p+\tilde{\mu})^2+\Delta^2},\nonumber
\end{equation}
\begin{equation}\label{Spectra-2}
\varepsilon'=\pm \sqrt{(p-\tilde{\mu})^2+4\Delta^2,}\quad
 \overline{\varepsilon}'=\pm \sqrt{(p+\tilde{\mu})^2+4\Delta^2}.
\end{equation}

At this point, we observe that in the high-density region required for the existence of the CFL phase the leading parameter is the quark chemical potential $\mu$, whose contribution to the thermodynamic potential occurs only through diagrams with internal lines of fermions. This allows to integrate in the ghost fields and neglect the ghost contribution to the gluon polarization tensors.  In a similar way, we can neglect all the gluon self-interactions in (\ref{GeffectiveZ}) and (\ref{GammaExpansion}). Moreover, the leading contribution in the infinity sum in (\ref{GammaExpansion}) comes from the term with the lowest order in the $\alpha_s$ expansion.  Therefore,  the gluon effective action reduces to 
\begin{equation} \label{Lagrangian-g}
 \int dx^4 [-\frac{1}{4}(\partial_{\mu}G_{\nu}^A-\partial_{\nu}G_{\mu}^A)^2-\frac{\xi^2}{2}(\partial_{\mu}G_{\mu}^A)^2+\frac{1}{2}G^A_{\mu}\Pi^{\mu\nu}_{AB}G^B_{\nu}],
\end{equation}
on which we used a covariant gauge $F_\xi=\xi\partial^\mu G_{\mu}^A=0$, with $\xi$ the gauge fixing parameter. 

The tensor $\Pi^{\mu\nu}_{AB}$ in (\ref{Lagrangian-g}) represents the fermion contribution to the one-loop gluon self-energy. In the hard-loop approximation \cite{Bellac} its leading part is \cite{Rischke2000}
\begin{eqnarray}\label{PO}
\Pi_{\mu \nu}^{AB}(p_0=0, \textbf{p}\rightarrow 0) = 
[\hat{\tilde{m}}_D^2 \delta_{\mu0}\delta_{\nu0} +\hat{\tilde{m}}_M^2\delta_{\mu
i}\delta_{\nu i}]\delta^{AB}, \end{eqnarray}
with the Debye and Meissner masses defined in Eq. (\ref{Masses}) with the replacement $\mu \rightarrow \tilde{\mu}$. Here, we introduced the notations $\hat{\tilde{m}}_D=\tilde{m}_D \theta({\Delta}-p)+\sqrt{3}\tilde{m}_g\theta(\tilde{\mu}-p)\theta(p-\Delta)$ and $\hat{\tilde{m}}_M=\tilde{m}_M\theta(\Delta-p) $ to take into account the limitation to soft external momenta of  $\Pi^{\mu\nu}_{AB}(p)$ in the hard-loop approximation. Note that there is a different mass value depending on the momentum interval. For the Debye mass there are two values, $\tilde{m}_D$ corresponding to the interval $0<p<\Delta$, which is the infrared region of color superconductivity,  and $\sqrt{3}\tilde{m}_g$ corresponding to $\Delta<p<\tilde{\mu}$, which is in the infrared domain of dense matter in the normal phase, where the Debye mass takes the known value $\sqrt{3}\tilde{m}_g$ \cite{Bellac, Rischke}. We should note that, while the Debye mass is different from zero once there is finite density, the Meissner mass is only confined to the color superconducting domain $p<\Delta$. 

With the help of (\ref{Lagrangian-g}), we can now integrate the gluons in (\ref{GeffectiveZ}) to find
\begin{equation} \label{OmegaG}
  \Omega_{G}=-\frac{8}{2\beta}\sum \hspace{-0.49cm}\int \frac{d^{4}p^E}{\left( 2\pi
\right) ^{3}}\ln det[i\Delta_{\mu\nu}^{-1}(p,\hat{\tilde{m}}_M,\hat{\tilde{m}}_D)],
\end{equation}
 with gluon inverse propagator
  \begin{equation} \label{Delta}
 i\Delta_{\mu\nu}^{-1}(p,\hat{\tilde{m}}_M,\hat{\tilde{m}}_D)=p^2\delta_{\mu\nu}+\hat{\tilde{m}}_D^2\delta_{\mu0}\delta_{\nu0}+\hat{\tilde{m}}_M^2\delta_{\mu i}\delta_{\nu j}+(1-\xi^2)p_{\mu}p_\nu
\end{equation}
and compact notation
 \begin{equation}\label{Matsubara}
\sum_{\it l}\hspace{-0.47cm}\int \frac{d^{4}p^E}{\left( 2\pi\right) ^{3}}=\frac{-1}{i\beta}\sum_{p_4}\int \frac{d^3p}{(2\pi)^3},
\end{equation}
for the three-momentum integral and sum in the Matsubara frequencies $p_4=\frac{2n\pi}{\beta},  n=0,\pm1,\pm2,...$ The factor 8 in the right-hand side of (\ref{OmegaG}) accounts for the degeneracy in the gluon spectra. 

Taking the Feynman gauge ($\xi=1$) and performing the Matsubara sum we obtain
\begin{equation}\label{TP-gluons}
 \Omega_{G}(T)=\frac{2}{\pi^2}\int_0^\Lambda dp p^2\left[\sqrt{p^2+\hat{\tilde{m}}_D^2}+3\sqrt{p^2+\hat{\tilde{m}}_M^2}+\beta^{-1}\ln(1-e^{\beta\sqrt{p^2+\hat{\tilde{m}}_D^2}})+3\beta^{-1}\ln(1-e^{\beta\sqrt{p^2+\hat{\tilde{m}}_M^2}})\right],
 \end{equation}

Therefore, in the $T\to 0$ limit, the gluon part of the thermodynamic potential becomes
\begin{equation}\label{TP-gluons-T0}
 \Omega_{g}=\Omega_{G}(T=0)=\frac{2}{\pi^2}\int_0^\Lambda dp p^2\left (\sqrt{p^2+ \tilde{m}^2_D \theta({\Delta}-p)+3\tilde{m}^2_g\theta(\tilde{\mu}-p)\theta(p-\Delta)}+3\sqrt{p^2+\tilde{m}_M^2\theta(\Delta-p)}\right )
 \end{equation}

The thermodynamic potential of a gauged-NJL model has been also investigated in a recent paper \cite{Huang}. However, this work did not consider color superconductivity and the study was done in the presence of a magnetic field.

Combining all the contributions, we obtain that the zero-temperature, mean-field thermodynamic potential of model (\ref{Lagrangian}) in the hard-loop approximation is given by
\begin{equation}\label{modelo}
\Omega=\Omega_{q}+ \Omega_{g}  - \Omega_{vac}
\end{equation}
where we subtracted the vacuum constant $\Omega_{vac}\equiv \Omega(\mu=0, \Delta=0)$.

The dynamical quantities $\Delta$ and  
$\rho$ should be found from the equations

\begin{equation} \label{Gap-Eq}
\frac{\partial\Omega}{\partial\Delta}=0,\;\;\;\;\frac{\partial\Omega_q}{\partial\tilde{\mu}}=0
\end{equation}

We call attention to two well-known but worth noticing features associated with the vector interactions of the model.  First, as the mean value $\langle\bar{\psi}\gamma_0\psi\rangle$ enters in the covariant derivative as a shift to the particle chemical potential, the effective chemical potential for the baryon charge is now $\tilde{\mu}$ instead of $\mu$. Second, while the solution of the gap equation (first equation in (\ref{Gap-Eq})) is a minimum of the thermodynamic potential, the solution of the second equation is a maximum \cite{Kitazawa2002}, since it defines, as usual in statistics, the particle number density $\rho$.

Finally, let us define the model parameters that will be used in the numerical calculations of next sections. Following a standard procedure, we define the energy cutoff $\Lambda=602.3$ MeV and the quark-antiquark coupling $G_S\Lambda^2=1.835$ and $K \Lambda^5=12.36$ to fit $f_\pi$, $m_\pi$, $m_K$ and $m_{\eta'}$ to their empirical values in the sharp cutoff regularization \cite {Rehberg}. Then, the diquark coupling $G_D$, that produces a gap $\Delta\simeq10$ MeV at $\mu=500$ MeV, is found to be $G_D=1.2 G_S$. A similar ratio $G_D/G_S$ was already considered  in \cite{GD-GS} to investigate the M-R relationship in hybrid compact stars with color superconducting cores. 
Changing $\Lambda$ in a few percentage, while simultaneously modifying $G_D$ to produce the same value of $\Delta$, does not affect our qualitative results. As for the values of the vector coupling, it is known that if the vector channel is originated from a Fierz transformation of a local color current-current interaction, the resulting coupling strength is $G_{V}=0.5 G_S$. If instead, one starts from the molecular instanton liquid model or the PNJL model, the Fierz transformations give rise to values of $G_V$ much more smaller \cite{GV-Vacuum}. Based on these considerations, $G_V$ is usually taken as a free parameter in the range $G_V=(0-0.5)G_S$. 

\section{Gluons Effects on the EOS of the CFL phase of Strange Stars}

In this section we shall explore the effects of the gluon rest-energy on the EOS of the CFL matter and the implications for strange stars. All our discussion will be based on the zero-temperature limit since compact stars typically have $\mu\gg T$ and our ultimate goal will be to investigate the star structure, which is insensitive to small temperature effects.

The pressure $P$ and energy density $\epsilon$ can be found from the thermodynamic potential (\ref{modelo}) as
\begin{equation}\label{energy}
P= -(\Omega_{q}+ \Omega_{g}  - \Omega_{vac})+(B-B_0), \quad \epsilon = \Omega_{q}+ \Omega_{g}  - \Omega_{vac} + \tilde{\mu} \rho-(B-B_0)
\end{equation}
Notice that the chemical potential that multiplies the particle number density $\rho$ in the energy density is $\tilde{\mu}$ instead of $\mu$. This result can be derived following the same calculations of Ref. \cite{Israel} to find the quantum-statistical average of the energy-momentum tensor component $\tau_{00}$, using for the present case the mean-field Lagrangian (\ref{MF-Lagrangian}). 

In (\ref{energy}), we added, as usual, the bag constant B ($B_0$ is introduced to ensure that $\epsilon=P=0$ in vacuum). In the MIT bag model, $B$ was introduced as a phenomenological input parameter to account for the free energy cost of quark matter relative to the confined vacuum \cite{Bag-Const}. Nevertheless, in the NJL model, the bag constant can be calculated in the mean-field approximation as a dynamical quantity related to the spontaneous breaking of chiral symmetry \cite{Oertel}.
Following this dynamical approach, we have that the bag constant is given in our model by
\begin{widetext}
\begin{eqnarray}\label{Dynamical-B}
B=\sum_{i=u,d,s}\left[\frac{3}{\pi^2}\int_0^\Lambda
p^2dp\left(\sqrt{m_i^2+p^2}-\sqrt{p^2}\right)
-2G_S \langle \overline{\psi}_i \psi_i \rangle \right]+4K\langle\overline{\psi}_u \psi_u \rangle\langle\overline{\psi}_d \psi_d \rangle\langle\overline{\psi}_s \psi_s \rangle
\end{eqnarray}
\end{widetext}
where  $m_i$ is the dynamical quark masses, the quark condensates for the different flavors are given by
\begin{equation}
\langle\overline{\psi}_i \psi_i \rangle=-\frac{3}{\pi^2}\int_{p_{Fi}}^\Lambda p^2 dp \frac{m_i}{\sqrt{m_i^2+p^2}},
\label{Quark-Condensates}
\end{equation}
with $p_{Fi}=(\pi^2\rho_i)^{1/3}$ being the Fermi momenta depending on the densities of each flavor $\rho_i=\langle \psi_i^\dag \psi_i\rangle$; and the dynamical masses are found from the gap equations
\begin{equation}
m_i=4G_S\frac{3}{\pi^2}\int_{p_{Fi}}^\Lambda p^2 dp \frac{m_i}{\sqrt{m_i^2+p^2}}+2K\frac{9}{\pi^4}\int_{p_{Fi}}^\Lambda p^2 dp \frac{m_j}{\sqrt{m_j^2+p^2}}\int_{p_{Fi}}^\Lambda p^2 dp \frac{m_k}{\sqrt{m_k^2+p^2}}
\label{Gap-Eq}
\end{equation}
where $i\neq j \neq k$ take values in the flavor set ${u,d,s}$.

The vacuum bag constant in (\ref{energy}) is given by
\begin{equation} 
B_0=B|_{\rho_u=\rho_d=\rho_s=0}
\end{equation}

For the densities we are considering in the CFL phase ($\rho>\rho_0$), the phase transition to the restored chiral phase (with $m_i=0$) already took place. Hence, in our case $B=0$ and the only remaining bag parameter in (\ref{energy}) is $B_0$, given by 
\begin{widetext}
\begin{eqnarray}\label{Dynamical-B0}
B_0=\frac{9}{\pi^2}\left[\int_0^\Lambda
p^2dp \left (\sqrt{m^2+p^2}-\sqrt{p^2}+\frac{2G_Sm}{\sqrt{m^2+p^2}}\right ) \right ]-4K\left (\frac{3}{\pi^2} \right )^3\left [ \int_0^\Lambda dp p^2 \frac{m}{\sqrt{m^2+p^2}} \right ]^3
\end{eqnarray}
\end{widetext}
with the same vacuum dynamical mass for the three quarks found from    
\begin{equation}
1=4G_S\frac{3}{\pi^2}\int_{0}^\Lambda p^2 dp \frac{1}{\sqrt{m^2+p^2}}+2K\frac{9}{\pi^4}\left [\int_{0}^\Lambda p^2 dp \frac{m}{\sqrt{m^2+p^2}}\right ]^2
\label{Gap-Eq-Vac}
\end{equation}
For the parameter set under consideration one obtains $B_0=57.3$ MeV/fm$^3$ \cite{Oertel}.

We call attention that due to the lack of confinement in the NJL model, the contribution of the confined vacuum to the pressure and energy density cannot be consistently derived in the framework of this model, what always result in certain degree of ambiguity in the proposed value for the bag constant.

\subsection{Absolute stability}

The first thing we need to elucidate is what is the range of parameters that ensures absolute stability of the quark matter phase here considered. Absolute stability of the strange matter exists as long as the matter energy per baryon number at zero pressure 
\begin{equation}\label{E/A}
E/A|_{P=0} = \frac{(\Omega_q-\Omega_{vac}^{q}+B_0)|_{P=0} + \tilde{\mu}_0 \rho}{\rho/3}=\frac{-\Omega_g+\Omega_{vac}^{g} + \tilde{\mu}_0 \rho}{\rho/3},
\end{equation}
remains below the corresponding value for the iron nucleus (roughly 930 MeV). Here $\tilde{\mu}_0$ is the value of the chemical potential at zero pressure. When the parameters are such that $E/A<930 MeV$, the deconfinement of nuclear matter into this phase of strange matter is energetically favored. 
In (\ref{E/A}) we took into account the relation between the baryon and quark number densities $\rho_B=\frac{1}{3}(\rho_u+\rho_d+\rho_s)=\frac{1}{3}\rho$. 

\begin{figure}
\begin{center}
\includegraphics[width=0.6\textwidth]{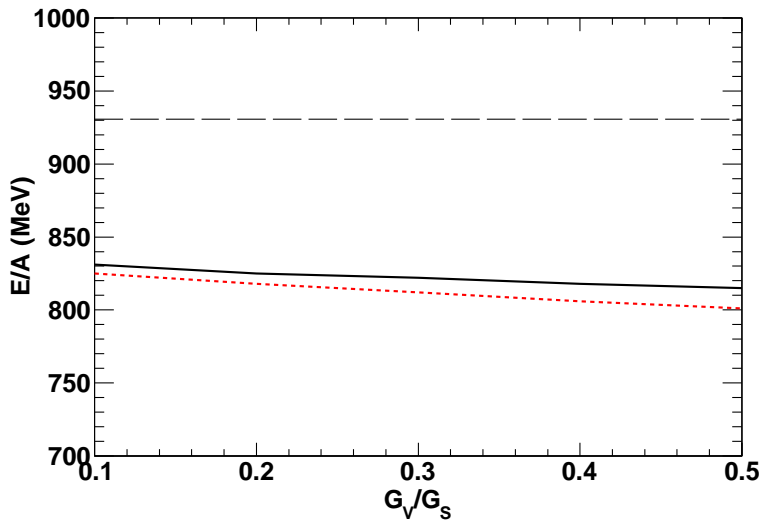}
\caption{(Color online) Energy per baryon number of quark matter at zero pressure as a function of the vector coupling. The horizontal long dashed line indicates the limiting value of 930 MeV (energy per baryon number of the iron nucleus). The solid line is for CFL matter without gluons, and the dashed line, for CFL matter with gluons.} \label{figStab}
\end{center}
\end{figure}

Fig. \ref{figStab} shows the behavior of $E/A|_{P=0}$ with the vector coupling $G_V$ for the CFL matter with and without gluons. The variation of the energy per baryon with the vector interaction at zero pressure is very smooth. This is expected because of the pressure normalization at low density, so the stability of the CFL matter is little affected by $G_V$. The effect of gluons on the energy per baryon number at zero pressure will depend on how much it lowers the pressure at a given density. If the effect were big enough to shift the zero pressure point to high values of quark density (high $\mu$), the last term in Eq. (\ref{E/A}) would push the energy up and destabilize the system. Nevertheless, the presence of gluons (dashed line) shifts the zero pressure point to a higher value of the density but not high enough as to take the system out of the absolute stability region. This shift being small, the zero pressure point is now closer to the minimum that usually exists in the energy vs. density curve, leading to a slightly lower energy per baryon number than without the gluons.

\subsection{EOS of strange quark matter with gluon contribution}

\begin{figure}
\begin{center}
\includegraphics[width=0.49\textwidth]{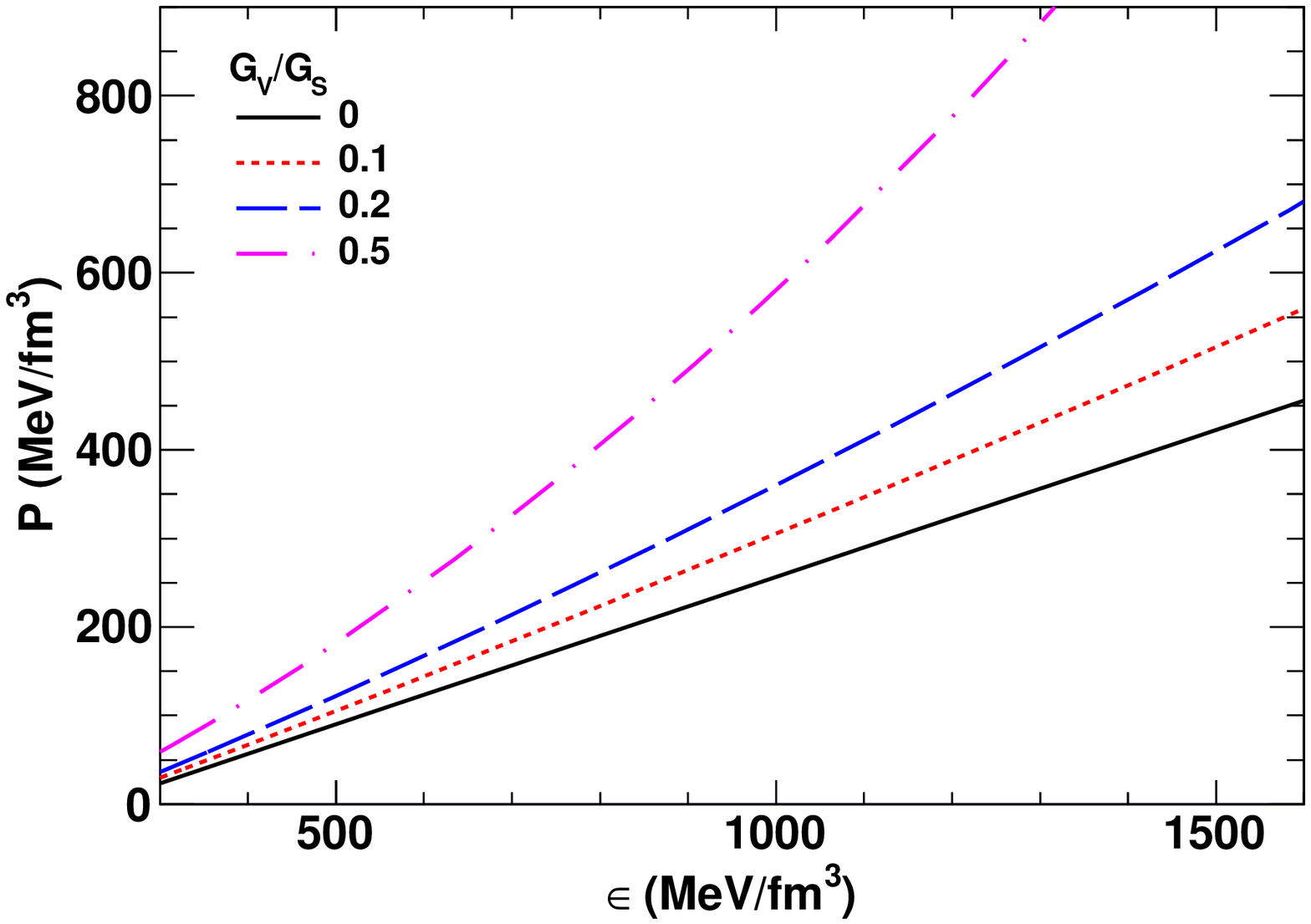}
\includegraphics[width=0.49\textwidth]{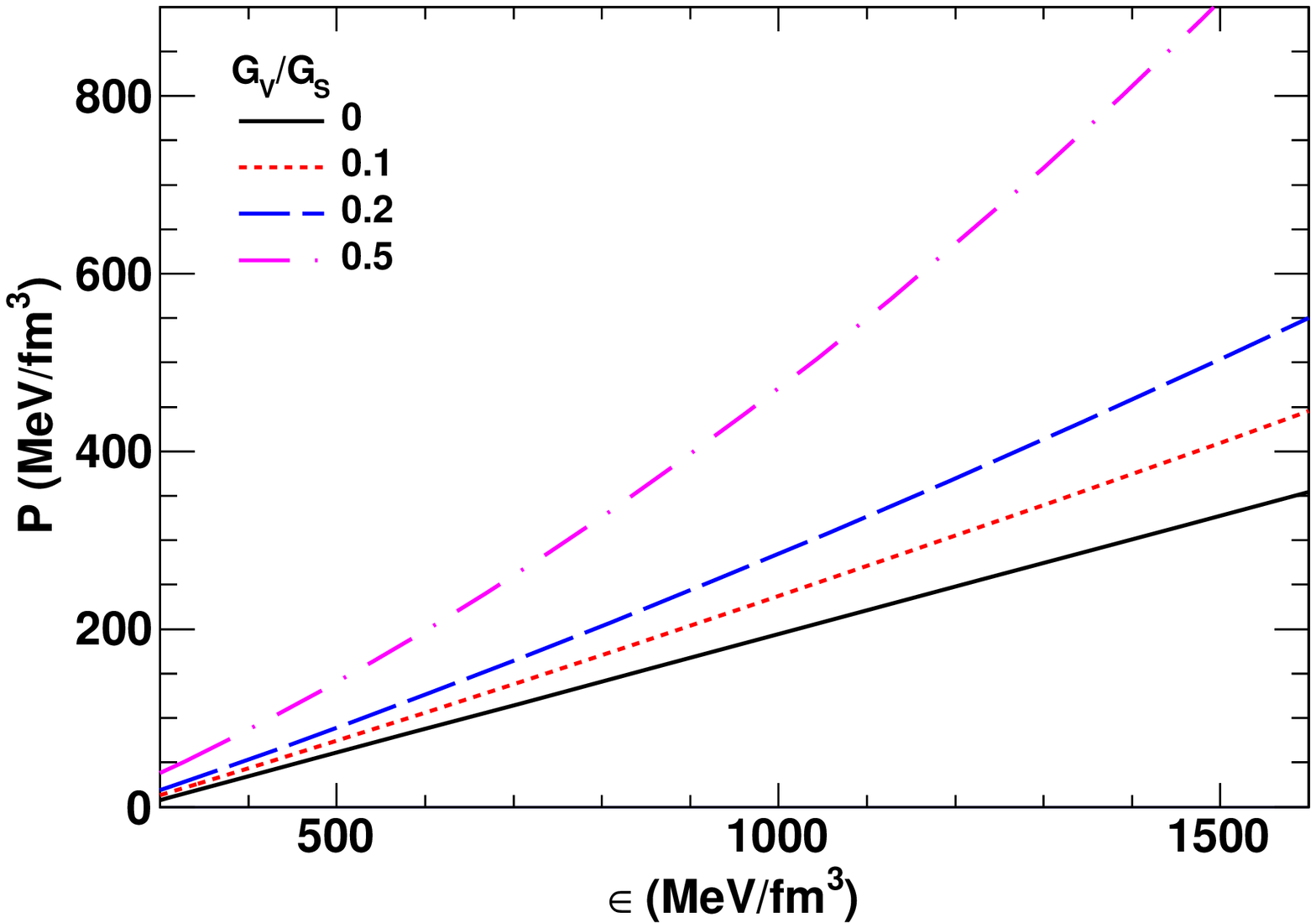}
\caption{(Color online) Equation of state for CFL matter with gluon contribution (right panel) and without it (left panel), for different values of the vector coupling $G_V$.}\label{figEoS}
\end{center}
\end{figure}

Fig. \ref{figEoS} displays the EOS for a range of $G_V$ with and without the gluons contribution to the thermodynamic potential. The inclusion of gluons degrees of freedom soften the EOS, as can be seen by comparing the two panels. Only relatively high values of $G_V$ can make the EOS of the CFL matter with gluons stiffer than the EOS of the regular CFL matter with no gluons and no vector coupling, at least in the energy density region relevant for compact stars. Of course, if we compare the curve for the CFL matter at certain value of $G_V$ with the corresponding one including gluons and the same coupling value, the former will be stiffer than the second one, independent of the $G_V$ value.

We can also note how the shift in pressure due to a change in $G_V$ for CFL matter with gluons for the same value of energy density increases with the increase of the energy density. While at energy densities smaller than $500$ MeV/fm$^3$ there is a negligible change even for $G_V/G_S=0.5$, at $\epsilon=1500$ MeV/fm$^3$, the jump in pressure is really noticeable.  This is an expected result, since the effect of the vector interaction is proportional to the baryon density and should be more significant in the high density region.

\subsection{Speed of sound in strange quark matter}

Another parameter that characterizes the dense matter is the speed of sound $v_S^2=dP/d\epsilon$, which measures how fast a disturbance in the pressure travels in the dense medium. It can be used to constraint the EOS, since it has to obey some natural conditions like causality $v_S\le c$, and thermodynamic stability $v_S^2 >0$.  A nonrelativistic gas of nuclear matter has a very small speed of sound. The opposite extreme, an ultrarelativistic gas of free particles, has $v^2_S/c^2=\frac{1}{3}$, and adding perturbative interactions or masses just leads to $v^2_S/c^2<\frac{1}{3}$ \cite{ Kurkela,
PRD81}. Based on these facts, it has been speculated that for the range of intermediate densities meaningful for neutron stars, the speed of sound should lie somewhere between the values of these two extremes. These facts have led to conjecture the existence of a fundamental bound $v^2_S/c^2\le\frac{1}{3}$ for the speed of sound of strongly interacting quark matter systems. However, as has been recently argued \cite{PRL114}, this bound seems hard to be reconciled with the observation of neutron stars with $2M_{\odot}$, at least for hybrid stars with all the reasonable low density EOS.  
\begin{figure}
\begin{center}
\includegraphics[width=0.49\textwidth]{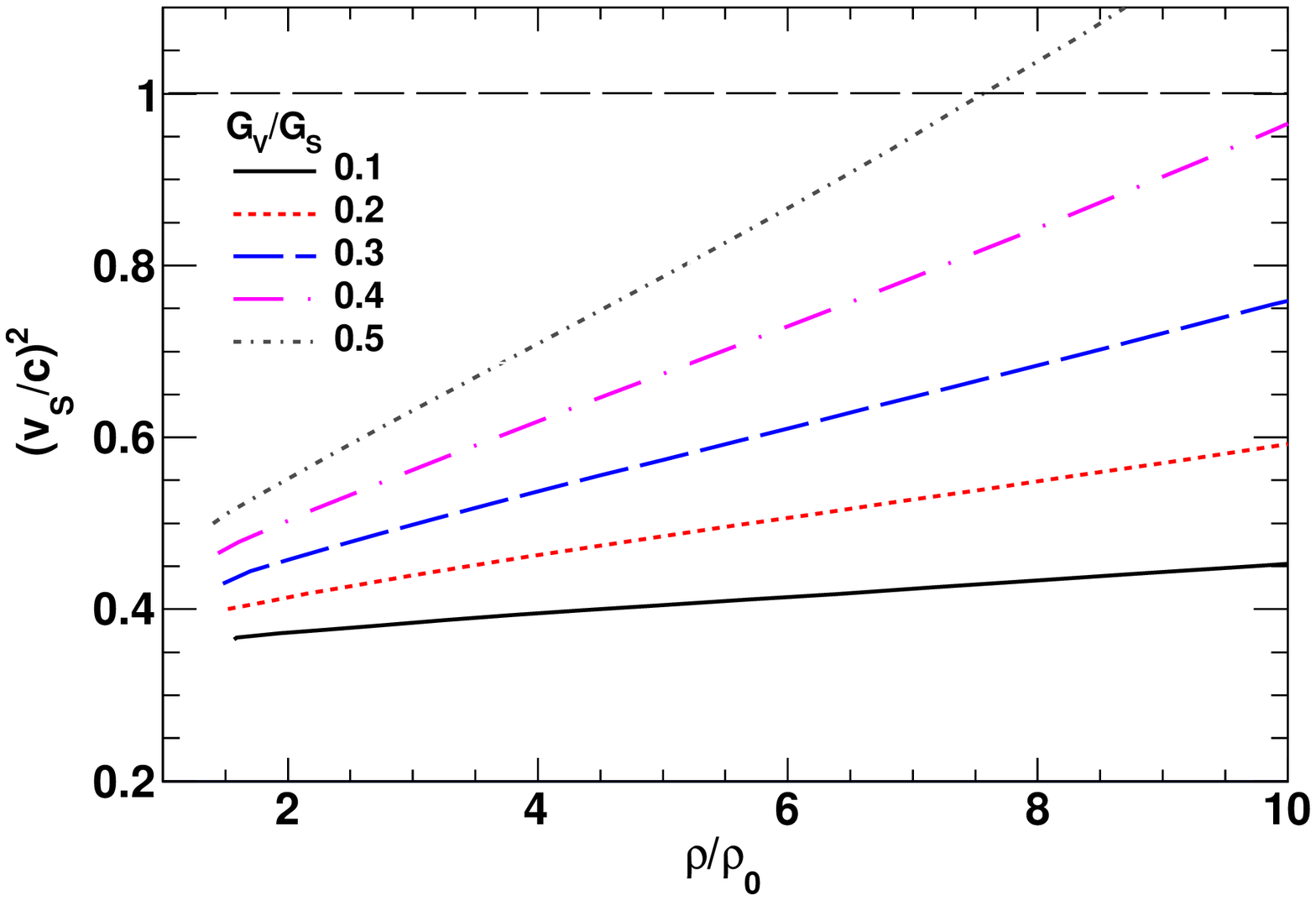}
\includegraphics[width=0.49\textwidth]{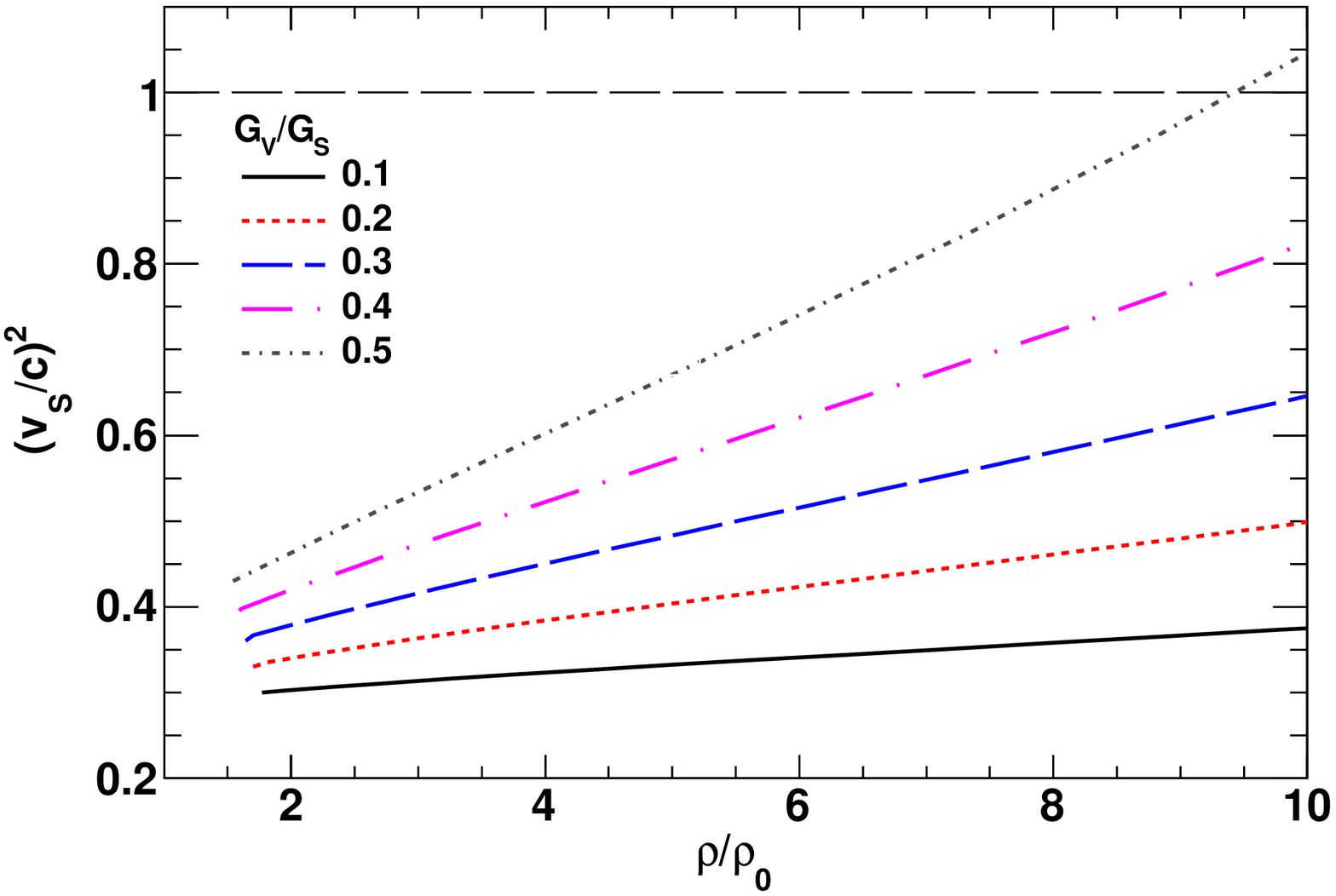}
\caption{(Color online) Sound speed as a function of the baryon density for CFL matter without the gluon term (left panel) and with it (right panel), for different values of the vector coupling strengths. The starting point to the left of each curve corresponds to the zero pressure point. The horizontal long dashed line determines the causality limit ($v_S=c$).}\label{soundsp}
\end{center}
\end{figure}

In Fig. \ref{soundsp} the sound speed is plotted as a function of the baryon density for CFL matter without gluons (left panel) and with gluons (right panel). The sound speed is very sensitive to the presence of the vector interactions, as can be seen from the slopes of the curves in two panels of Fig. \ref{soundsp}. Since the gluons have Debye and Meissner masses in the CFL phase, their presence tends to decrease the speed of sound for each given density and coupling $G_V$.  As expected, a vector interaction tends to increase the speed since it hardens the EOS. 
Notice that despite the softening of the EOS by the gluons, the condition $v^2_S/c^2<\frac{1}{3}$ is violated for all the vector interactions compatible with the absolute stability.

\section{Gluon Effects on the M-R relationship of strange matter in the CFL phase}

An immediate application of the EOS  just derived
is to use it to construct stellar models of compact stars. In this regard,
there are two distinct possibilities: the formation of strange stars with the strange quark matter in the CFL phase described by the EOS (\ref{energy}), where the parameters are chosen to ensure absolute stability; or the formation of hybrid stars, on which the quark phase can be metastable, so the star cannot be all made of quark matter but it can exist in a core that is surrounded by nuclear matter. Given the results of Fig. \ref{figStab} we are going to investigate the first possibility. With this aim, we should obtain the M-R sequence for strange stars made of matter with EOS given by (\ref{energy}) and explore how the gluons and vector interactions of the model can affect the sequence. Of particular relevance is to determine whether this sequence can reach the observed mass values of the compact objects, PSR J1614-2230 and PSR J0348+0432 with $M=1.97\pm 0.04M_{\odot}$ \cite{Demorest} and $M=2.01\pm 0.04M_{\odot}$ \cite{Antoniadis} respectively.

\begin{figure}
\begin{center}
\includegraphics[width=0.49\textwidth]{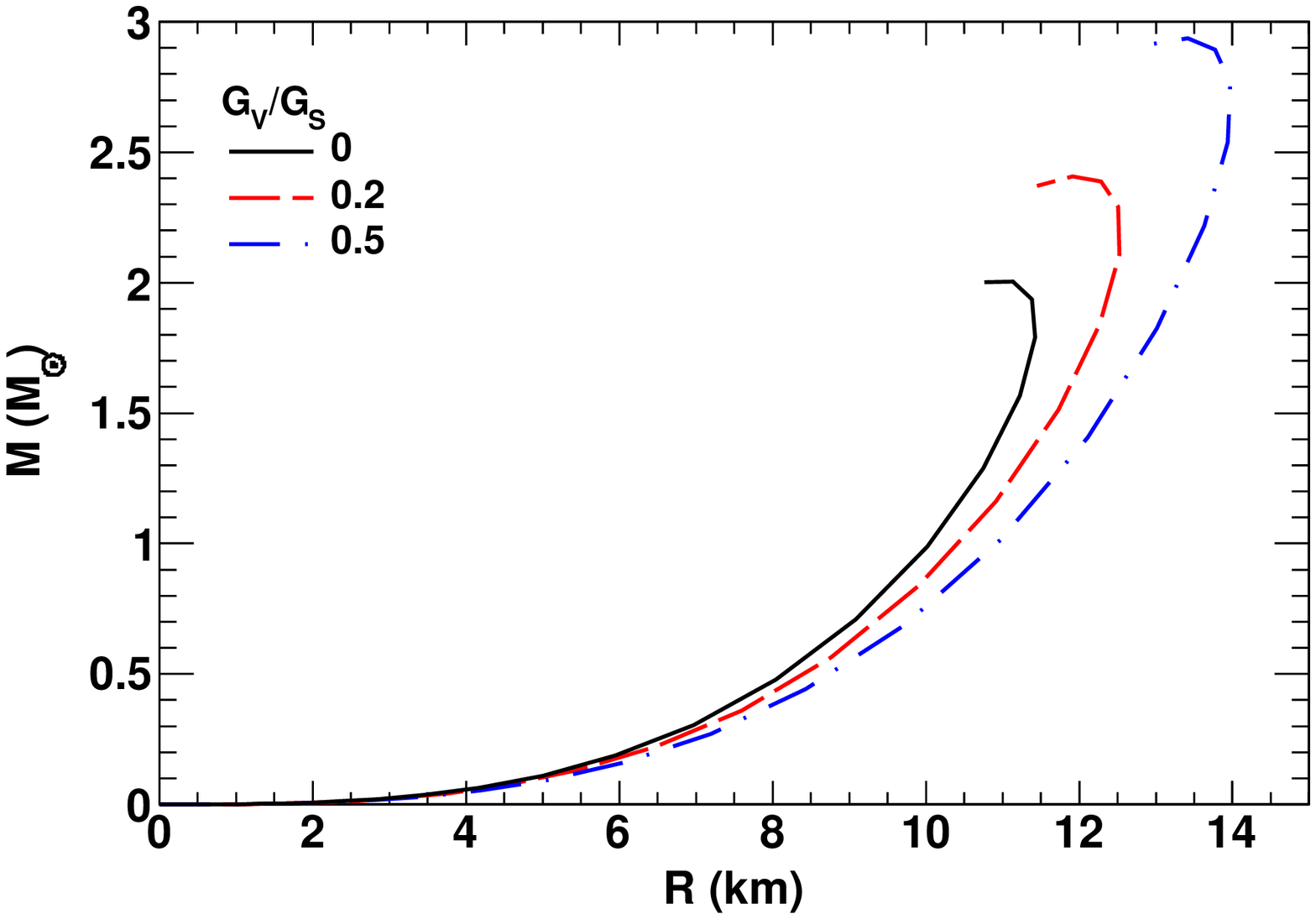}
\includegraphics[width=0.49\textwidth]{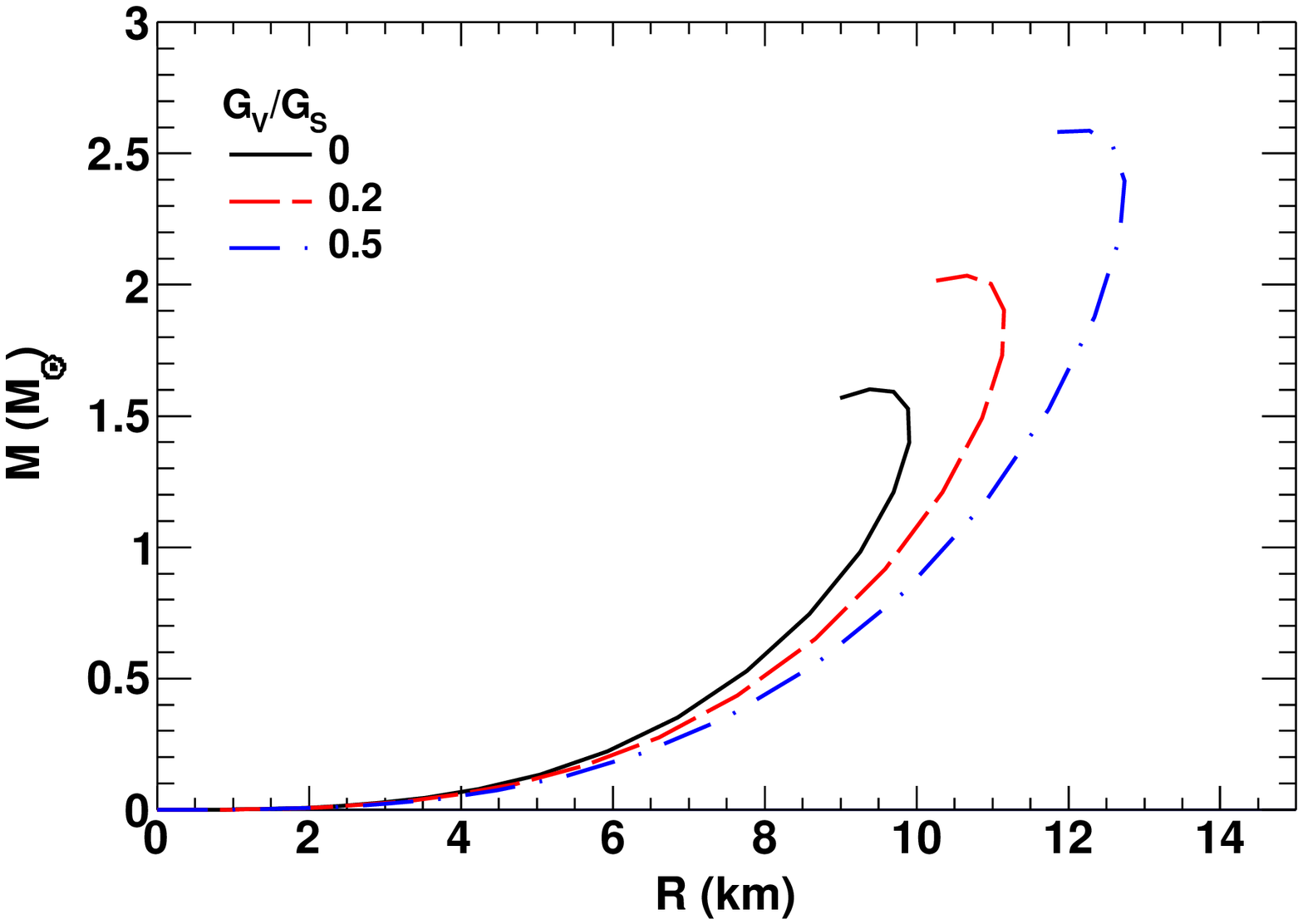}
\caption{(Color online) Mass-radius relationship of strange stars made of CFL matter with (right panel) and without (left panel) the gluon contribution. The softening of the EOS by the gluons significantly decreases the maximum mass of the sequence, which for vector interaction with $G_V/G_S<0.2$ remains below 2$M_\odot$.} \label{figMR}
\end{center}
\end{figure}

The M-R relationship is obtained by integrating the relativistic
equations for stellar structure, that is, the well-known Tolman-Oppenheimer-Volkoff (TOV) and mass continuity equations, which in  natural units, $c = G = 1$ are given by
\begin{eqnarray}
\frac{dM}{dR}&=&4\pi R^2\epsilon \label{TOV1}\\
\frac{dP}{dR}&=&-\frac{\epsilon M}{R^2}\Big(1+\frac{P}{\epsilon}\Big)\Big
(1+\frac{4\pi R^3P}{M}\Big)\Big(1-\frac{2M}{R}\Big)^{-1}\label{TOV2}
\end{eqnarray}
with $P$ and $\epsilon$ taken from (\ref{energy}). The corresponding M-R sequences are shown in Fig. \ref{figMR} for EOS of quark matter with and without the gluon contribution. Comparing them, it is evident that the gluons decrease the maximum mass for each sequence up to $20 \%$, this effect being more prominent for lower values of the vector interaction. Sequences including gluons within our model cannot reach $2M_{\odot}$ if $G_V/G_S<0.2$. Here, we should point out that several results actually suggest that a low vector coupling between quarks may be favored at high-densities  \cite{GV-Vacuum}. Even more, as discussed in \cite{Kashiwa}, the vector interaction makes the chiral phase transition weaker in the low $T$ and high $\mu$ region, with the possibility that  the transition becomes a crossover in the region
when the interaction is strong enough \cite{Kitazawa2002, Abuki}, what is not expected from Lattice calculations. Then, the absence of the vector interaction would be preferable in the high density region. The same conclusion was reached by studying the phase transition from the hadronic phase to the quark phase in Ref. \cite{Bentz}. There, it was found that the hadron-quark phase transition may take place only at a small $G_V/G_S$ ratio. For a larger ratio, the repulsive vector interaction makes the NJL phase too stiff to allow the crossing from the hadronic phase. However, such a phase transition could take place if one assumes zero vector interactions in the quark matter side and further considering effects that soften the EOS like diquark condensation.  

Although at present there is no definitive way to determine the value of the $G_V$ coupling in the high-dense region, all these results provide an indication that $G_V$ could be significantly small in the quark phase. If this is the indisputable case, then if $G_V/G_S<0.2$, strange stars composed entirely of CFL quark matter with gluons, within the formalism presented here, should be disregarded, as they cannot explain the mass measurements of the mentioned compact stars PSR J1614-2230 and PSR J0348+0432.

\section{Conclusions}

In this paper we investigate the EOS of strange stars described by a three-flavor gauged-NJL model with vector and diquark channels. In the region of large densities and within the mean-field approximation, the ground state of the system is the same as in the CFL phase, but with the replacement of the baryon chemical potential by an effective chemical potential $\widetilde{\mu}=\mu-G_{V}\rho$. In addition, the EOS of the system contains the contribution of the rest energy of the gluons, which acquires Debye and Meissner masses in the color superconducting background through polarization operators with lines of fermions in the hard-loop approximation. 
The pressure and energy density defining the EOS were then used to calculate the M-R relationship of strange stars in such a parameter range.

 We have found that gluons soften the EOS of the CFL matter. This renders maximum stellar masses lower than the observed $\sim 2M_{\odot}$ if $G_V/G_S<0.2$. This effect could exclude pure color superconducting strange stars as possible candidates of compact stars unless vector interactions were strong enough at high densities. New interactions, as the diquark-diquark repulsion studied in Ref. \cite{Jason} or higher order quark interactions \cite{David}, could bring enough outward pressure to make the favored composition of neutron stars to be pure superconducting quark matter with low vector interaction. Other quark-matter phases as 2SC with strong diquark coupling and including gluon contributions would have to be investigated as possible candidates for hybrid stars.

We underline that while the energy eigenvalue $\epsilon=0$ is always allowed in the partition function of non-relativistic bosons in a big box with periodic boundary conditions, so at zero temperature these particles always occupy the zero energy ground state with zero momentum, and hence zero pressure, things are very different for relativistic bosons. In the relativistic case, if the bosons have nonzero rest energy, they produce a positive contribution to the energy density and a negative term in the pressure. Since gluons in the color superconductor acquire rest energy due to their density-dependent Debye and Meissner masses (\ref{Masses}), these energies affect the EOS of the CFL matter at $T=0$, thereby giving rise to all the interesting effects discussed in this paper.

An important point to be indicated is that in the framework of the NJL model, the only vacuum contribution that can be calculated from first principles is that corresponding to the chiral condensation, as presented in Sec. III, while the additional contribution that will be related to confinement cannot be derived in a self consistent way. This is a limitation of the phenomenological NJL-type models used to describe the color superconducting stellar medium.

To finish, we want to stress once again that due to the limitations of the used model, which is expressed in terms of a set of undetermined parameters ($\Lambda$, B, $G_V$, etc.), any statement regarding the existence or not of strange stars has to be taken
more as an indication than as a definitive conclusion.   

\begin{acknowledgments}
The work of EJF and VI has been supported by DOE Nuclear Theory grant DE-SC0002179. LP acknowledges the financial support received from the Brazilian funding agencies CNPq, Conselho Nacional de Desenvolvimento Cient\'ifico e Tecnol\'ogico, and Fapesp, Funda\c c\~ao de Amparo \`a Pesquisa do Estado de S\~ao Paulo (2013/26258-4), and the hospitality of the UTEP Physics Department where this work was conducted. 

\end{acknowledgments}

\end{document}